\documentclass[ draft
  ]
  {aipproc}

\layoutstyle{6x9}

\usepackage[fleqn]{amsmath}
\usepackage{bbm}
\usepackage{graphicx}
\usepackage{epsfig}

\def\pl#1#2#3{Phys.~Lett.~{\bf B{#1}}, #3 ({#2})}
\def\np#1#2#3{Nucl.~Phys.~{\bf B{#1}}, #3 ({#2})}
\def\prl#1#2#3{Phys.~Rev.~Lett.~{\bf #1}, #3 ({#2})}
\def\pr#1#2#3{Phys.~Rev.~{\bf D{#1}}, #3 ({#2})}

\begin{document}
{\normalsize \hspace{-0.55cm}
DESY-07-186\hfill\mbox{}\\
October 2007\hfill\mbox{}\\}
\vspace{-1cm}
\title[ ]{Baryogenesis -- 40 Years Later\footnote{13th International
Symposium on Particles, Strings and Cosmology, Imperial College
London, July 2007}}
\date{}
\classification{98.80.Cq,95.35.+d,12.60.Jv,95.30.Cq}
\keywords      {Baryogenesis, Leptogenesis, Dark Matter}

\author{Wilfried Buchm\"uller}
{address={Deutsches Elektronen-Synchrotron DESY, Notkestrasse 85, 
22603 Hamburg, Germany} }



\begin{abstract}
The classical picture of GUT baryogenesis has been strongly modified by
theoretical progress concerning two nonperturbative features of the
standard model: the phase diagram of the electroweak theory, and baryon
and lepton number changing sphaleron processes in the high-temperature  
symmetric phase of the standard model. We briefly review three viable
models, electroweak baryogenesis, the Affleck-Dine mechanism and leptogenesis
and discuss the prospects to falsify them. All models are closely tied
to the nature of dark matter, especially in supersymmetric theories.
In the near future results from LHC and gamma-ray astronomy will shed new 
light on the origin of the matter-antimatter asymmetry of the universe.

\end{abstract}

\maketitle


\section{Matter-Antimatter Asymmetry}

The cosmological matter-antimatter asymmetry can be dynamically generated if 
the particle interactions and the cosmological evolution satisfy
Sakharov's conditions \cite{sak67},
\begin{itemize}
\item baryon number violation,
\item $C$ and $C\!P$ violation,
\item deviation from thermal equilibrium.
\end{itemize}
Although the baryon asymmetry is just a single number, it provides an
important connection between particle physics and cosmology. In his seminal 
paper, 40 years ago, Sakharov not only stated the necessary conditions
for baryogenesis, he also proposed a specific model. The origin of the
baryon asymmetry were $C\!P$ violating decays of superheavy
`maximons' with mass $\mathcal{O}(M_\mathrm{P})$ at an initial temperature 
$T_i \sim M_\mathrm{P}$. The $C\!P$ violation in maximon decays was related
to the $C\!P$ violation observed in $K^0$-decays, and the violation 
of baryon number led to a proton lifetime $\tau_p > 10^{50}\ \mathrm{years}$,
much larger than current estimates in grand unified theories. 

\begin{figure}[t]
\includegraphics[height=6cm]{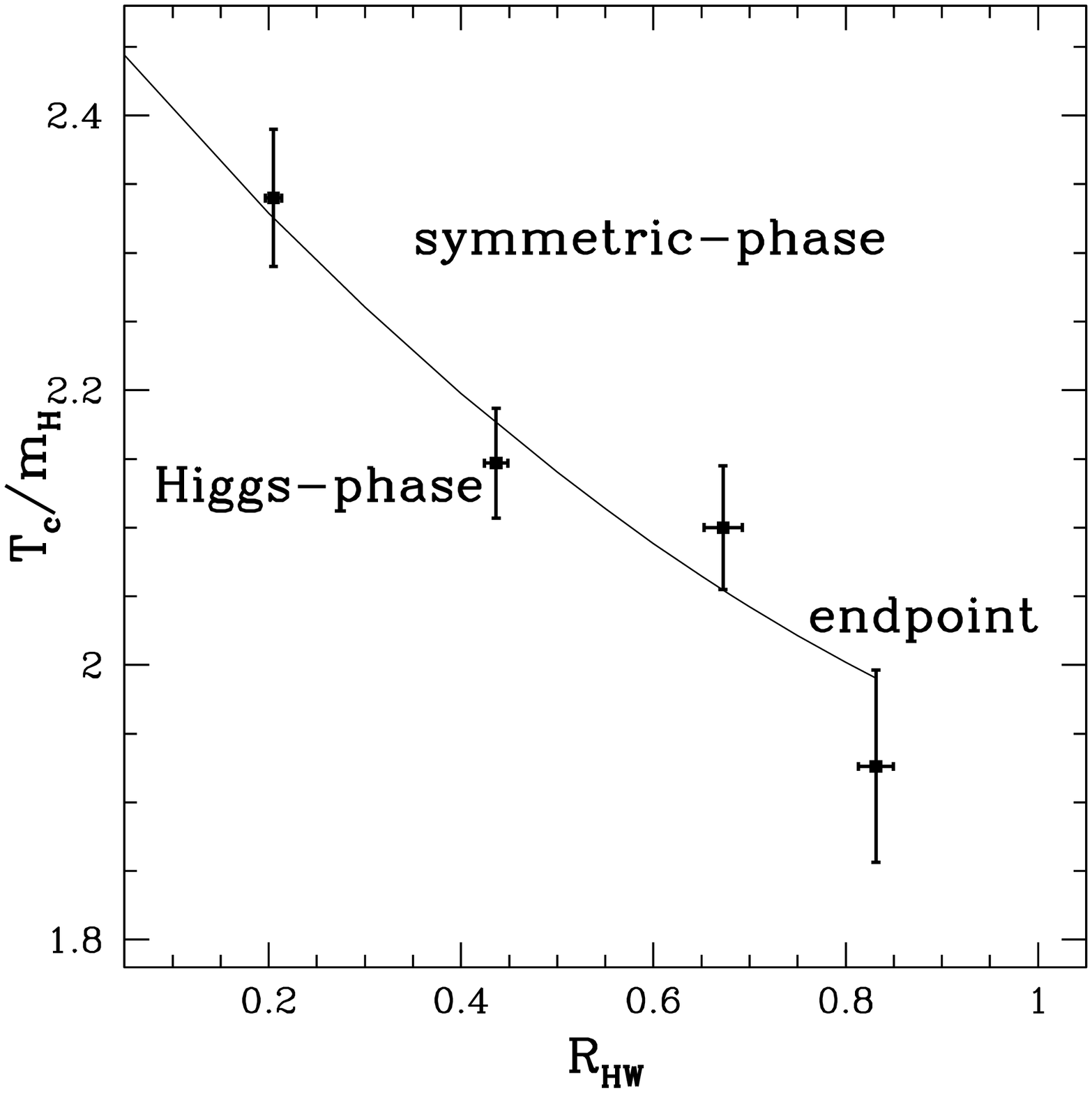}\hspace{1cm}
\includegraphics[height=6cm,width=8cm]{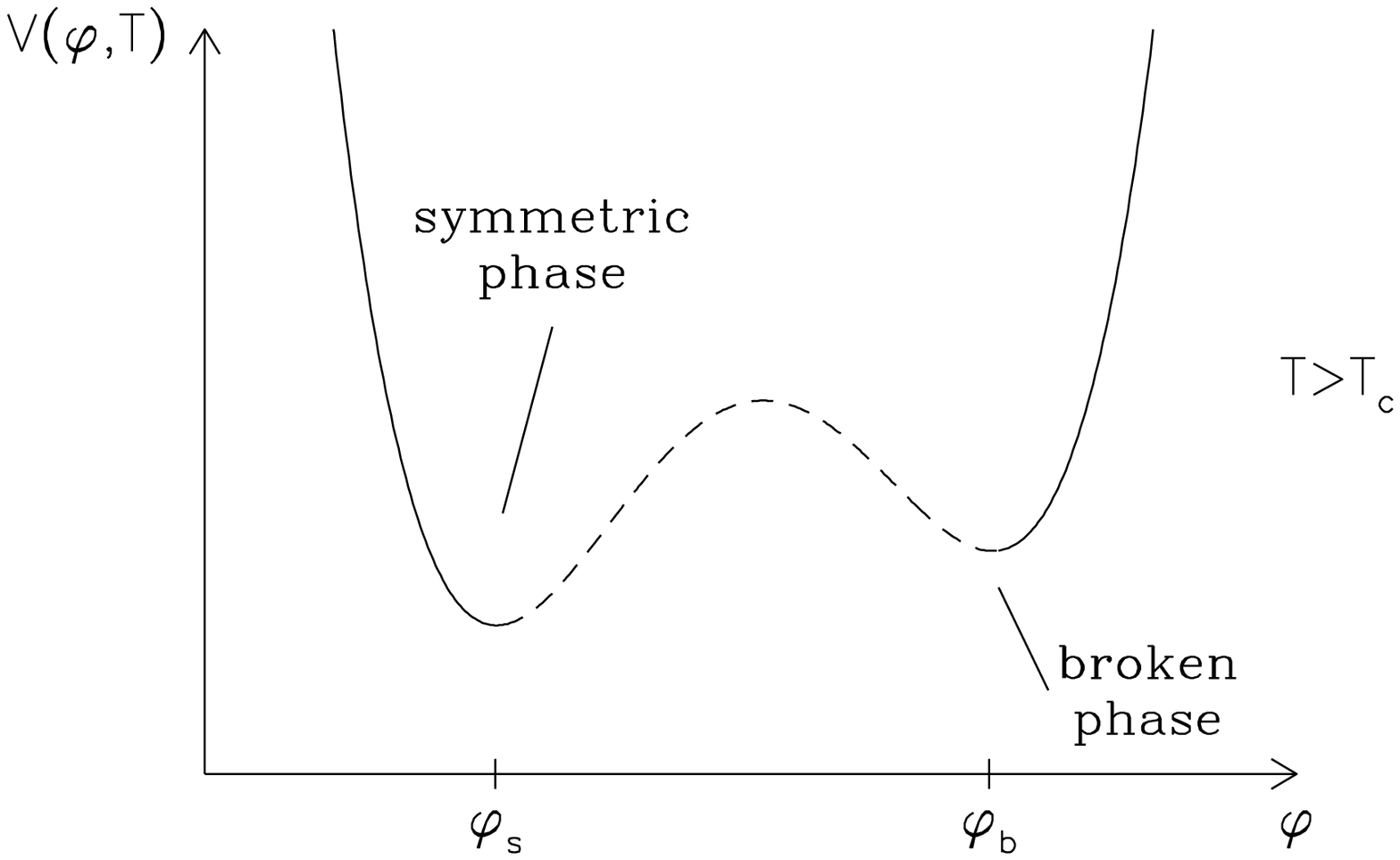}
\caption{{\it Left:} Critical temperature $T_c$ of the electroweak transition 
as function of $R_{HW}=m_H/m_W$; from \cite{cfh98}. {\it Right:} Effective
potential of the Higgs field $\varphi$ at temperature $T>T_c$.
\label{fig:ew}}
\end{figure}

At present there exist a number of viable scenarios for baryogenesis.
They can be classified according to the different ways in which Sakharov's
conditions are realized. In grand unified theories baryon number ($B$) and
lepton number ($L$) are broken by the interactions of gauge bosons and
leptoquarks. This is the basis of classical GUT baryogenesis (cf.~\cite{kt90}).
In a similar way, lepton number violating decays of heavy Majorana neutrinos
lead to leptogenesis \cite{fy86}. In the simplest version of
leptogenesis the initial abundance of the heavy neutrinos is generated
by thermal processes. Alternatively, heavy neutrinos may be produced in
inflaton decays or in the reheating process after inflation.
Because in the standard model baryon number, $C$ and $C\!P$ are not conserved, 
in principle the cosmological baryon asymmetry can also be generated at 
the electroweak phase transition \cite{krs85}. A further mechanism of 
baryogenesis can work in supersymmetric theories where the scalar potential has
approximately flat directions. Coherent oscillations of scalar fields can
then generate large asymmetries \cite{ad85}.

The theory of baryogenesis crucially depends on nonperturbative properties
of the standard model, first of all the nature of the electroweak transition.
A first-order phase transition yields a departure from thermal equilibrium. 
Fig.~1 shows the phase diagram of the electroweak theory, i.e. the critical
temperature in units of the Higgs mass, $T_c/m_H$, as function of the Higgs
mass in units of the W-boson mass, $R_{HW}=m_H/m_W$ \cite{cfh98,lr98}. For
small Higgs masses the phase transition is first-order; above a critical
Higgs mass, $m_H > m_H^c \simeq 72$~GeV, it turns into a smooth crossover 
\cite{bp94,klx96}. This upper bound for a first-order transition has to be 
compared 
with the lower bound from LEP, $m_H > 114$~GeV. Hence, there is no departure
from thermal equilibrium at the electroweak transition in the standard model.

The second crucial nonperturbative aspect of baryogenesis is the connection 
between baryon number and lepton number in the high-temperature, symmetric 
phase of the standard model. Due to the chiral nature of the weak interactions 
$B$ and $L$ are not conserved \cite{tho76}. At zero temperature this has no 
observable effect due to the smallness of the weak coupling. However, as the 
temperature reaches the critical temperature $T_c$ of the electroweak phase 
transition, $B$ and $L$ violating processes come into thermal 
equilibrium \cite{krs85}. The rate of these processes is
related to the free energy of sphaleron-type field configurations which carry
topological charge. In the standard model they lead to an effective
interaction of all left-handed fermions \cite{tho76} (cf.~Fig.~2), 
\begin{equation}
O_{B+L} = \prod_i \left(q_{Li} q_{Li} q_{Li} l_{Li}\right)\; ,
\end{equation}
which violates baryon and lepton number by three units, 
\begin{equation}
    \Delta B = \Delta L = 3\;. \label{sphal1}
\end{equation}
The sphaleron transition rate in the symmetric high-temperature phase
has been evaluated by combining an analytical resummation with numerical
lattice techniques \cite{bmr00}. The result is, in accord with previous 
estimates, that $B$ and $L$ violating processes are in thermal equilibrium for 
temperatures in the range
\begin{equation}
T_{EW} \sim 100\ \mbox{GeV} < T < T_{SPH} \sim 10^{12}\ \mbox{GeV}\;.
\end{equation}

\begin{figure}[t]
\includegraphics[height=6cm]{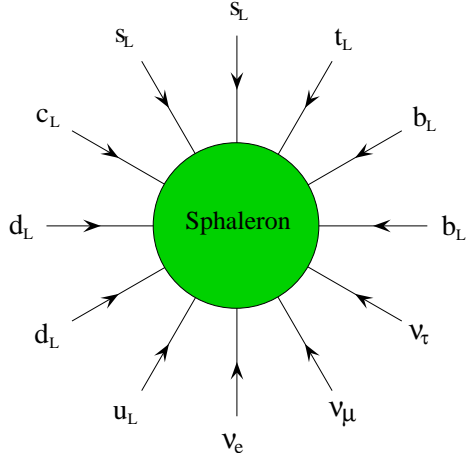}
\caption{One of the 12-fermion processes which are in thermal equilibrium
in the high-temperature phase of the standard model.
\label{fig:sphal}
}
\end{figure}

Sphaleron processes have a profound effect on the generation of the
cosmological baryon asymmetry.  
An analysis of the chemical potentials
of all particle species in the high-temperature phase yields the following
relation between the baryon asymmetry and the corresponding
$L$ and $B-L$ asymmetries,
\begin{equation}\label{basic}
\langle B\rangle_T = c_S \langle B-L\rangle_T = 
{c_S\over c_S-1} \langle L\rangle_T\;.
\end{equation}
Here $c_S$ is a number ${\cal O}(1)$. In the standard model with three 
generations and one Higgs doublet one has $c_s= 28/79$. 

We conclude that lepton number violation is necessary in order to 
generate a cosmological baryon asymmetry\footnote{In the case of Dirac 
neutrinos, which have extremely small Yukawa couplings, one can construct 
leptogenesis models where an asymmetry of lepton doublets is accompanied by an
asymmetry of right-handed neutrinos such that the total lepton
number is conserved and $\langle B-L \rangle_T = 0$
\cite{dlx00}.}. However, it can only be weak, 
because otherwise any baryon asymmetry would be washed out. The interplay of 
these conflicting conditions leads to important contraints on neutrino 
properties and on possible extensions of the standard model in general.   

\begin{figure}[t]
\includegraphics[height=6cm]{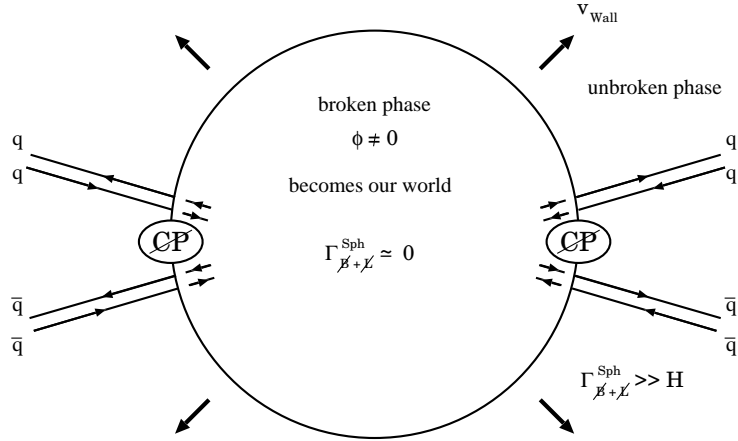}
\caption{Sketch of nonlocal electroweak baryogenesis. From \cite{ber02}.
}
\end{figure}

\section{Electroweak Baryogenesis}

A first-order electroweak phase transition proceeds via nucleation and growth
of bubbles (cf.~\cite{ber02,cli06}). This can provide the departure from 
thermal equilibrium, which is necessary for electroweak baryogenesis. $C\!P$
violating reflections and transmissions at the bubble surface then generate
an asymmetry in baryon number, and for a sufficiently strong phase transition 
this asymmetry is frozen in the true vacuum inside the bubble (cf.~Fig.~3).

\begin{figure}[b]
\includegraphics[height=6cm, width=7cm]{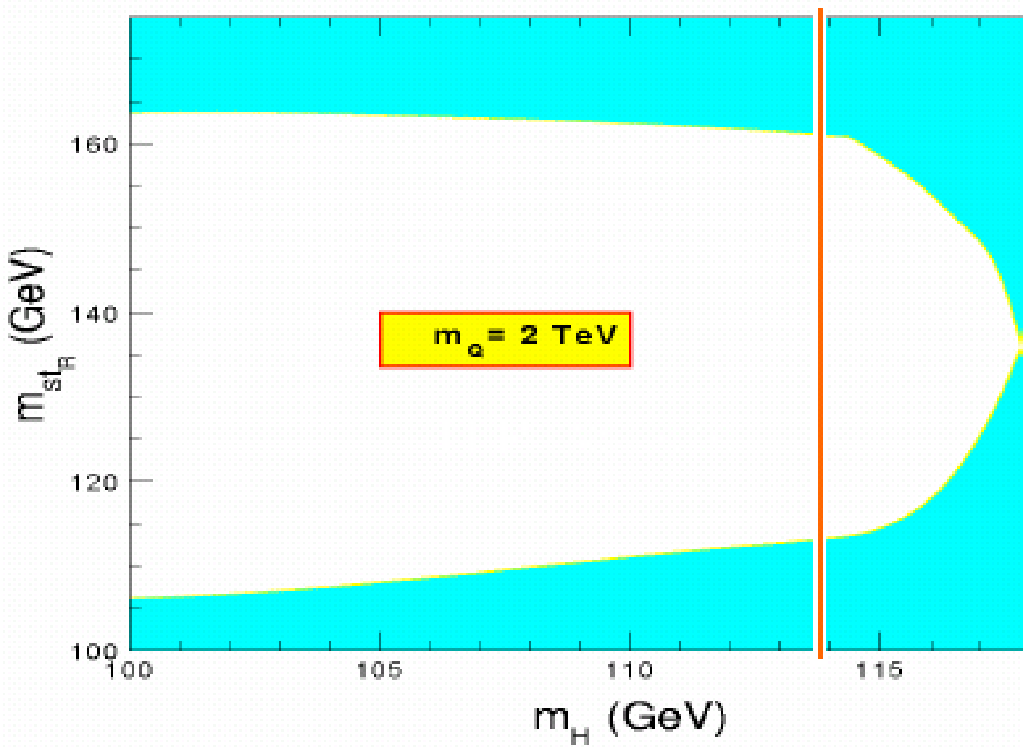}\hspace{1cm}
\includegraphics[height=6cm,width=7cm]{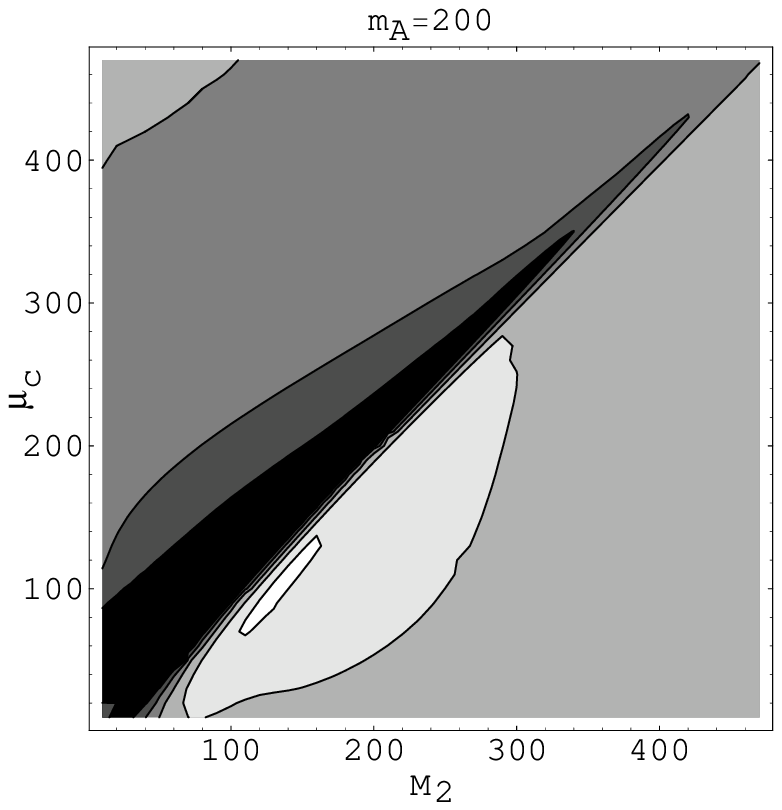}
\caption{{\it Left:} Upper and lower bounds on the scalar top mass $m_{st_R}$ 
as function of the Higgs mass $m_H$. From \cite{priv}. {\it Right:} In the 
black area of the ($\mu_c$,$M_2$) plane of $\mu$-parameter and gaugino mass
electroweak baryogenesis is viable. From \cite{hkx06}.
}
\end{figure}

As discussed in the previous section, in the standard model the electroweak
transition is just a smooth crossover. Hence, there is no departure from
thermal equilibrium and baryogenesis cannot take place. The situation changes
in two-Higgs doublet models (cf.~\cite{cli06,hub06}) and in supersymmetric 
extensions of the standard model where one can have a sufficiently strong 
first-order phase transition (cf.~\cite{cli06}). This requires, however,
a rather exceptional mass spectrum of superparticles. As the left panel of 
Fig.~4 shows, one scalar top-quark has to be lighter than the top-quark 
whereas other scalar quarks are 2~TeV heavy. Also gaugino masses have to be 
rather small (cf.~Fig.~4, right panel). 

Even more stringent constraints are obtained if the lightest neutralino is
required to be the dominant component of cold dark matter. This case has
been studied in detail for the nMSSM, a minimal extension of the MSSM with
a singlet field \cite{mmw04}. Fig.~5 shows the neutralino relic density as
function of the neutralino mass for various parameter sets of the model
represented by the scattered points. It is remarkable that the neutralino
has to be very light. This suggests that, should supersymmetry be
discovered at the LHC, the consistency of WIMP dark matter and electroweak
baryogenesis will be a highly non-trivial test of supersymmetric extensions
of the standard model. 

\begin{figure}[t]
\includegraphics[height=6cm]{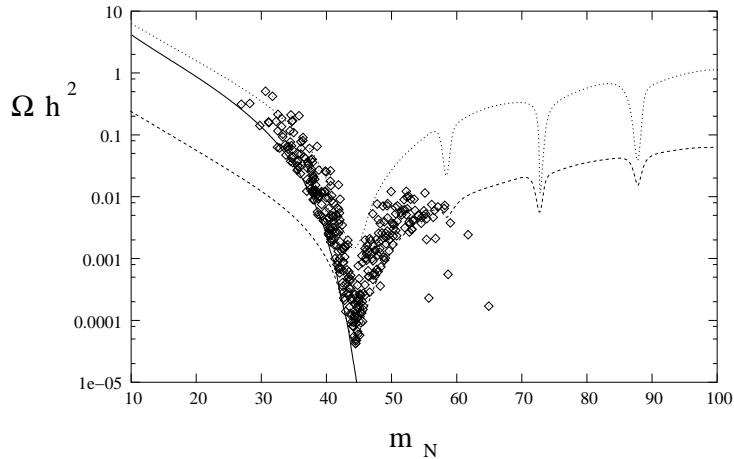}
\caption{Neutralino relic density as function of the neutralino mass in the
nMSSM for different parameter sets of the model (scattered points). From
\cite{mmw04}.
}
\end{figure}

\section{Affleck-Dine Baryogenesis}

In general the scalar potential of supersymmetric theories has many flat 
directions involving scalar fields which carry baryon or lepton number.
Typical examples in the MSSM are
\begin{equation}
(LH_u)\;,\quad (U^cD^cD^c)\;,
\end{equation}
where $L$, $H_u$, $U^c$ and $D^c$ denote lepton doublets, one of the Higgs 
doublets and quark fields, respectively.
During inflation these fields generically develop large vacuum expectation 
values. 
After inflation these condensates lead to coherent oscillations, 
which can store large baryon and lepton charge densities. The decay of these
condensates eventually converts the scalar charge densities to ordinary 
fermionic baryon and lepton number. 

This `AD mechanism' is a prominent example of nonthermal baryogenesis.
So far no `standard model' of AD baryogenesis has emerged, and it
appears difficult to falsify this scenario. On the other hand, certain
types of dark matter would strongly support the AD mechanism.  

\begin{figure}[t]
\includegraphics[height=6cm, width=7cm]{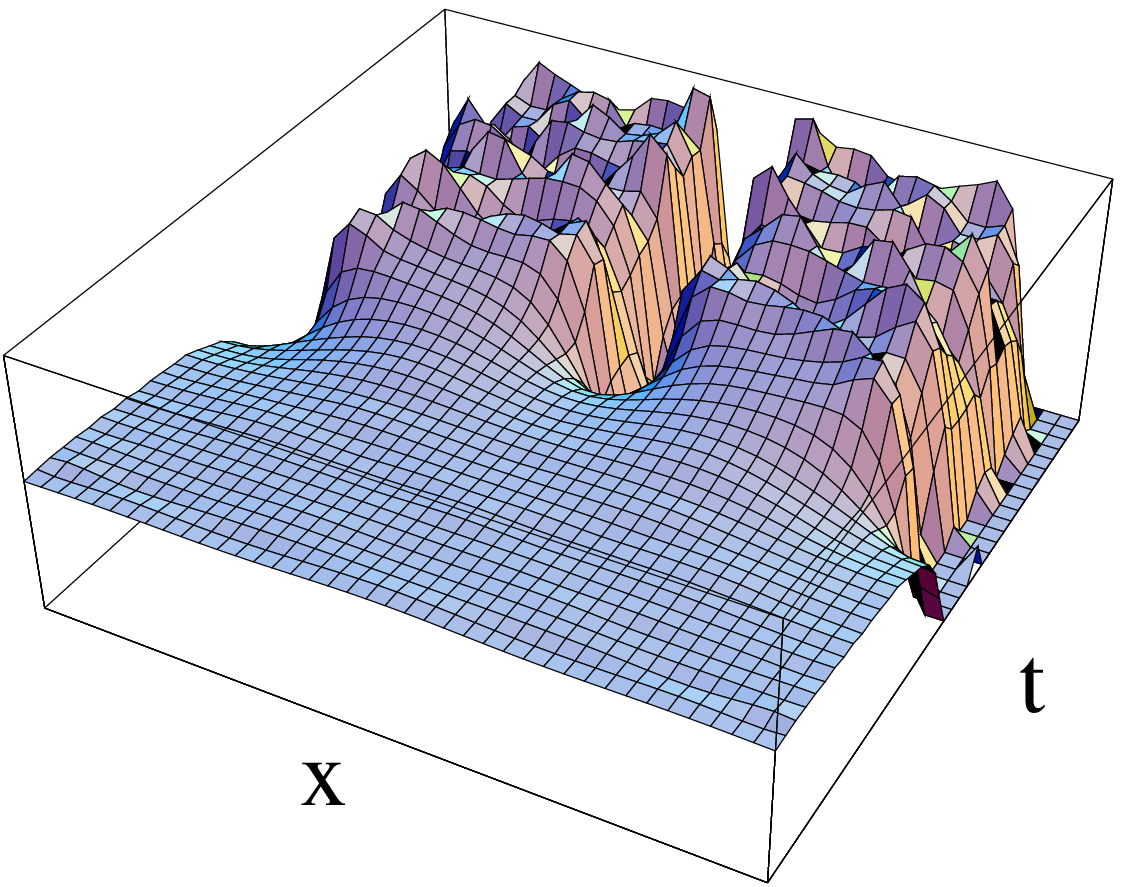}\hspace{1cm}
\includegraphics[height=5.8cm,width=6.5cm]{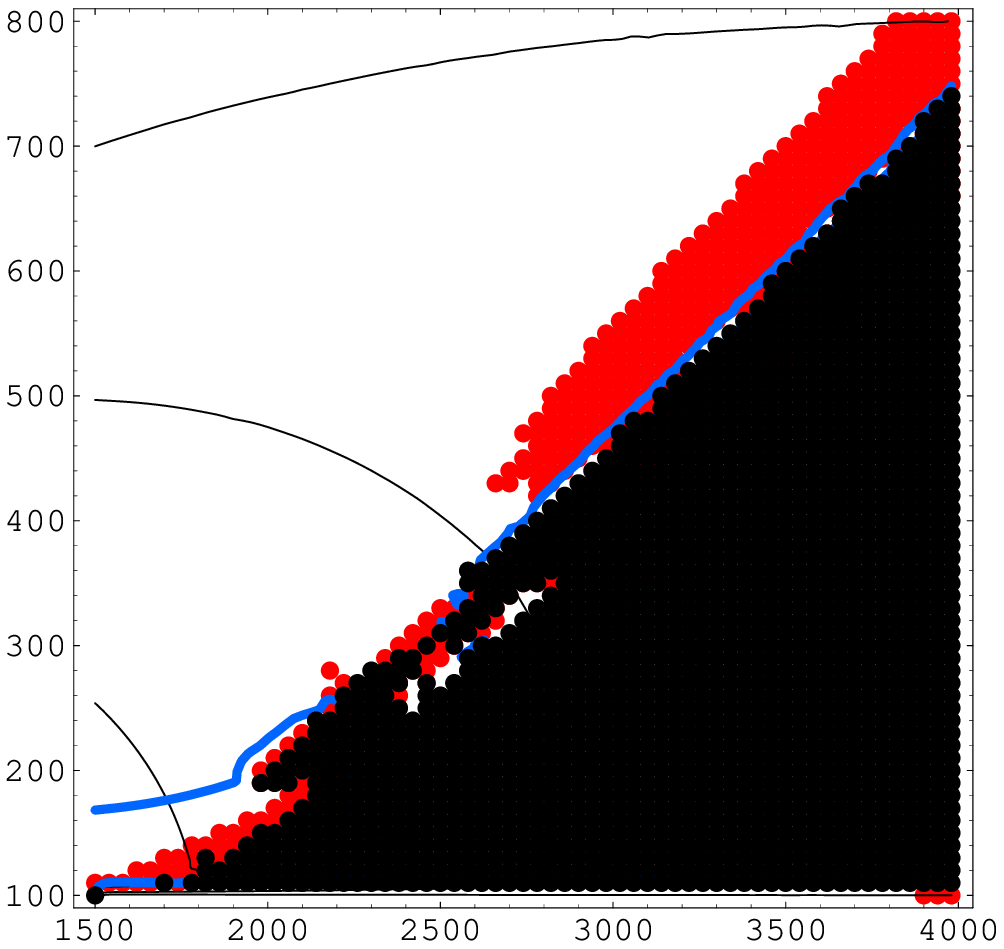}
\caption{{\it Left:} The charge density per comoving unit volume in (1+1)
dimensions for a sample potential during the period when the spatially
homogeneous condensate breaks up into high- and low-density domains which
are expected to form Q-balls. From \cite{dk03}. {\it Right:} Allowed domains
(red) of $m_0$~[GeV] (horizontal axis) and $M_{1/2}$~[GeV] (vertical axis) in
an mSUGRA model for nonthermally produced higgsino dark matter; the thin
black contours correspond to different Higgs masses. From \cite{fh02}. 
}
\end{figure}

In the case of the $U^cD^cD^c$ flat direction, the decay of the condensate can 
lead to the formation of Q-balls as illustrated in the left panel of Fig.~6 
(cf.~\cite{dk03}). These macroscopic objects with large baryon number and mass,
\begin{equation}
B_Q \sim 10^{26}\;, \quad M_Q \sim 10^{24}\ \mathrm{GeV}\;,
\end{equation}
can lead to striking signatures at Super-Kamiokande and ICECUBE.
Alternatively, the decay of unstable Q-balls can nonthermally produce
higgsinos which, for the parameters shown in Fig.~6, yield the observed
cold dark matter density. The identification of a neutralino LSP as
higgsino at LHC would be inconsistent with thermally produced WIMP dark 
matter. A discovery of higgsino dark matter in direct search experiments
could then be a hint for Q-balls as a possible nonthermal production
mechanism. In this way, as in the case of electroweak baryogenesis, the 
nature of
dark matter would provide a clue also for the origin of ordinary matter.

\section{Thermal Leptogenesis}

About 20 years ago, leptogenesis was suggested as the origin of matter
by Fukugita and Yanagida \cite{fy86}. The basis of this proposal is 
the seesaw mechanism which explains the smallness of the light neutrino masses 
by mixing with heavy Majorana neutrinos. The theory predicts six Majorana 
neutrinos as mass eigenstates, three heavy ($N$) and three light ($\nu$),
\begin{equation}
m_N \simeq M \, , \quad  m_\nu = - m_D^T{1\over M}m_D \;.
\label{seesaw}
\end{equation}
Here the Dirac neutrino mass matrix $m_D = h v$ is the product of the matrix 
$h$ of Yukawa couplings and the expectation value $v$ of the Higgs field 
$\phi$, which breaks the electroweak symmetry. If Yukawa couplings of the 
third generation are ${\cal O}(1)$, as it is the case for the top-quark, the
corresponding heavy and light neutrino masses are 
\begin{equation} 
M_3 \sim \Lambda_{GUT} \sim 10^{15}\ {\rm GeV}\ , \quad 
m_3 \sim \frac{v^2}{M_3} \sim\ 0.01\ {\rm eV}\;. 
\end{equation}
It is very remarkable that the light neutrino mass 
$m_3$ is of the same order as the mass differences $(\Delta m^2_{sol})^{1/2}$ 
and $(\Delta m^2_{atm})^{1/2}$ 
inferred from neutrino oscillations. This suggests that, via the seesaw 
mechanism, neutrino masses indeed probe the grand unification scale! 
The difference of the observed mixing patterns of quarks and leptons is a 
puzzle whose solution has to be provided by the correct GUT model. Like for 
quarks and charged leptons one expects a mass hierarchy also for the 
right-handed neutrinos. For instance, if their masses scale like the up-quark 
masses one has $M_1 \sim 10^{-5} M_3 \sim 10^{10}$~GeV.

The lightest of the heavy Majorana neutrinos, $N_1$, is ideally suited to 
generate the cosmological baryon asymmetry. Since it has no standard model 
gauge interactions it can naturally satisfy the out-of-equilibrium condition. 
$N_1$ decays to lepton-Higgs pairs then yield a lepton asymmetry 
$\langle L \rangle_T \neq 0$, which is partially converted to a baryon 
asymmetry $\langle B \rangle_T \neq 0$. The generated asymmetry is proportional
to the $C\!P$ asymmetry \cite{fps95} in $N_1$-decays which is conveniently 
expressed in the following form,
\begin{equation}
\varepsilon_1 = {\Gamma(N_1 \rightarrow l \phi) - 
 \Gamma(N_1 \rightarrow \bar{l} \bar{\phi})\over
 \Gamma(N_1 \rightarrow l \phi) + 
 \Gamma(N_1 \rightarrow \bar{l} \bar{\phi})} 
\ \simeq\  - {3\over 16\pi} {M_1\over (h h^\dagger)_{11} v^2}
 \mbox{Im}\left(h^* m_\nu h^\dagger\right)_{11}\;, \label{nice}
\end{equation}
where the seesaw mass relation (\ref{seesaw}) has been used. 
From the expression (\ref{nice}) one easily obtains a rough estimate for 
$\varepsilon_1$ in terms of neutrino masses. Assuming dominance of the 
largest eigenvalue of $m_\nu$, phases ${\cal O}(1)$ and 
approximate cancellation of Yukawa couplings in numerator and denominator one
finds,
\begin{equation}
\varepsilon_1\sim {3\over 16\pi}{M_1 m_3\over v^2}\sim\ 0.1\ {M_1\over M_3}\;,
\end{equation}
where we have again used the seesaw relation. Hence, the order of magnitude 
of the $C\!P$ asymmetry is approximately given by the mass hierarchy of the 
heavy Majorana neutrinos. For $M_1/M_3 \sim m_u/m_t \sim 10^{-5}$ one has 
$\varepsilon_1 \sim 10^{-6}$. 

Given the $C\!P$ asymmetry $\varepsilon_1$ one obtains for the baryon 
asymmetry,
\begin{equation}\label{basym}
\eta_B = {n_B - n_{\bar{B}}\over n_\gamma} 
       = -d \varepsilon_1 \kappa_f \sim 10^{-10}\;.
\end{equation}
Here the dilution factor $d \simeq 10^{-2}$ accounts for the increase 
of the number of photons in a comoving volume element between baryogenesis 
and today, and the efficiency factor $\kappa_f$ represents the effect of
washout processes in the plasma. In the estimate (\ref{basym}) we have assumed 
a typical value, $\kappa_f \sim 10^{-2}$. Thus the correct value of the baryon 
asymmetry is obtained as consequence of a large hierarchy of the heavy 
neutrino masses, which leads to a small $C\!P$ asymmetry, and the kinematical 
factors $d$ and $\kappa_f$ \cite{bp96}. 
The baryogenesis temperature,
\begin{equation}
T_B \sim M_1 \sim 10^{10}\ \mbox{GeV}\;,
\end{equation}
corresponds to the time $t_B \sim 10^{-26}$~s, which characterizes the next 
relevant epoch before electroweak transition, nucleosynthesis and 
recombination.

\begin{figure}[t]
\includegraphics[height=6.3cm, width=7cm]{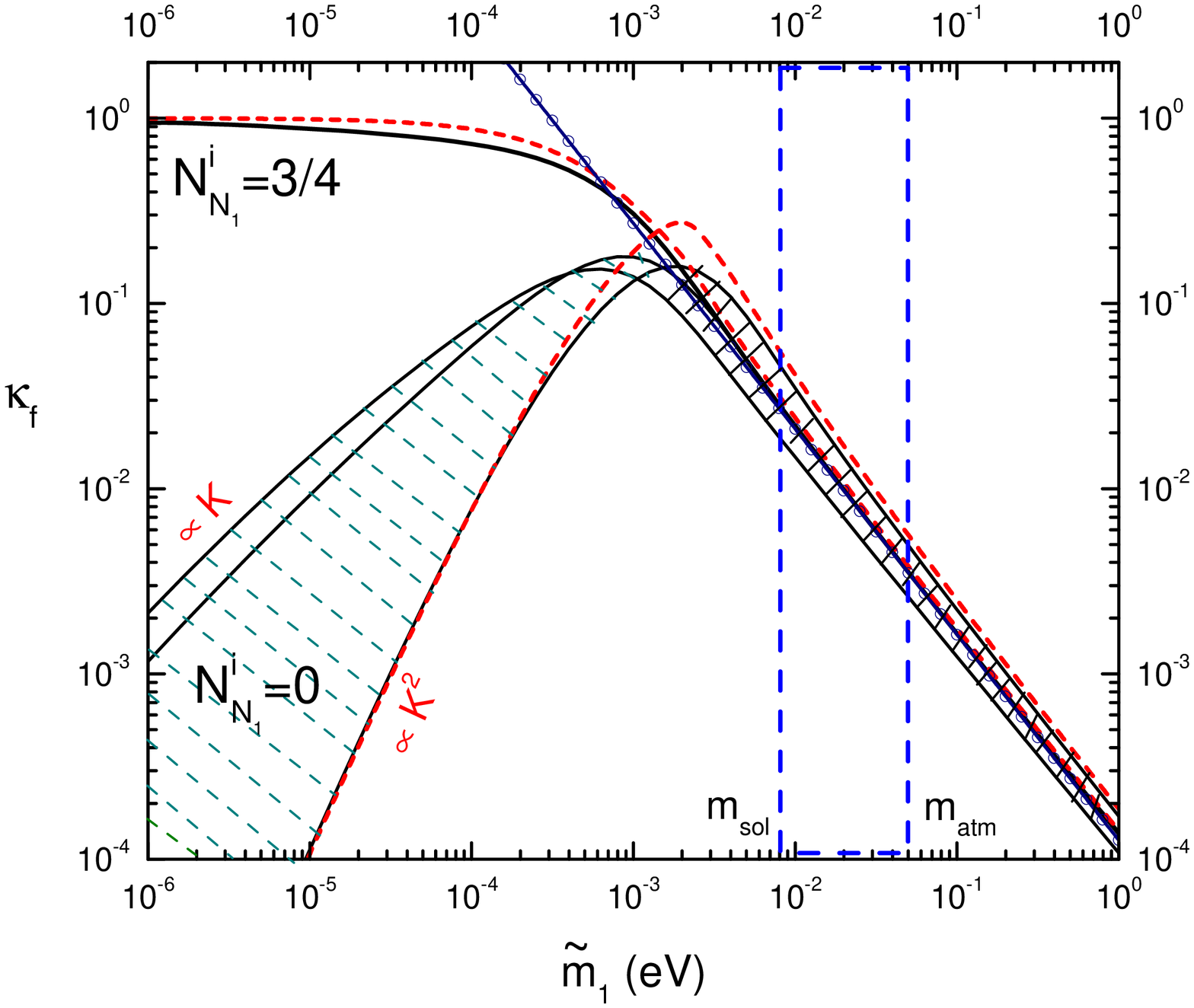}\hspace{1cm}
\includegraphics[height=6cm,width=6.5cm]{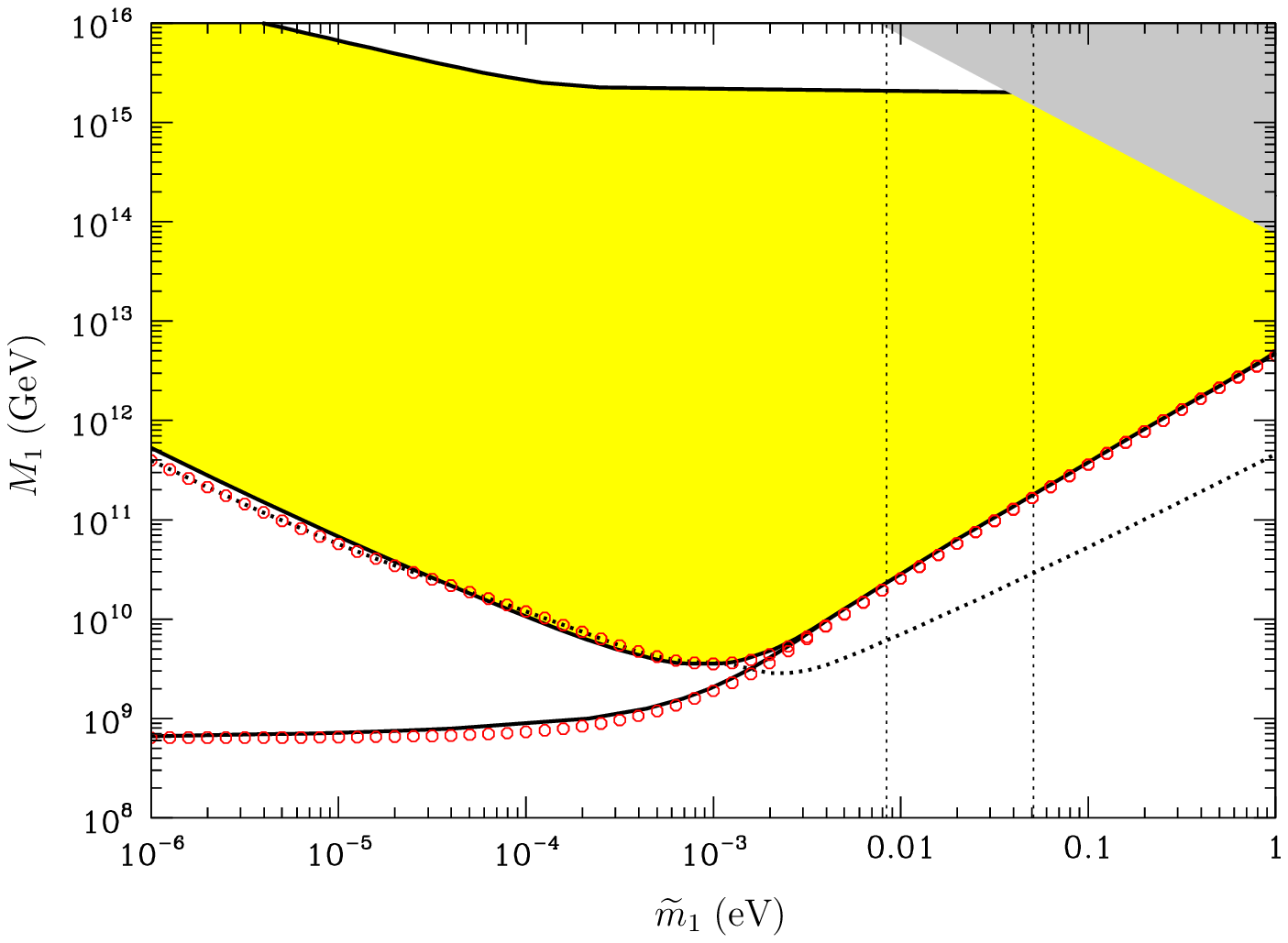}
\caption{{\it Left:} Final efficiency factor $\kappa_f$ as function of the 
effective neutrino mass $\widetilde{m}_1$. {\it Right:} Lower bounds on $M_1$
(analytical: circles) and the initial temperature $T_i$ (dotted line) as
functions of $\widetilde{m}_1$. Upper and lower curves correspond to zero
and thermal initial $N_1$ abundance, respectively. In both panels the vertical
dashed lines indicate the range ($m_{\rm sol}$,$m_{\rm atm}$). From 
\cite{bdp04}.
}
\end{figure}

During the past years the quantitative connection between thermal leptogenesis
and neutrino masses has been studied in great detail, in particular in the
simplest case of hierarchical heavy Majorana neutrinos. The crucial ingredients
are the upper bound on the $C\!P$ asymmetry $\varepsilon_1$ \cite{hmy02,di02}
and the analysis of the various production and washout processes in the
thermal plasma \cite{bdp02,bdp03,pu03,gnx03,bdp04}. One finds that successful
leptogenesis favours the light neutrino mass window \cite{bdp03}
\begin{equation}\label{win}
10^{-3}~{\rm eV} < m_i < 0.1~{\rm eV}\;.
\end{equation}
For $m_i > 10^{-3}$, the efficiency factor $\kappa_f$, and 
therefore the baryon asymmetry $\eta_B$, is independent of the initial $N_1$ 
abundance; furthermore, the final baryon asymmetry does not depend on the 
value of an initial baryon asymmetry generated by some other mechanism
(cf.~Fig.~7). 
Hence, the value of $\eta_B$ is entirely determined by neutrino properties.
For neutrino masses $m_i > 0.1\ {\rm eV}$ the $C\!P$ asymmetry $\varepsilon_1$
becomes too small and washout processes are too strong such that the generated
baryon asymmetry is too small. A second important result is a lower bound
on the baryogenesis temperature $T_B$ \cite{di02,bdp04} of about 
$10^9\ {\rm GeV}$, depending on $\widetilde{m}_1$ and the initial $N_1$
abundance. 

\begin{figure}[t]
\includegraphics[height=4cm]{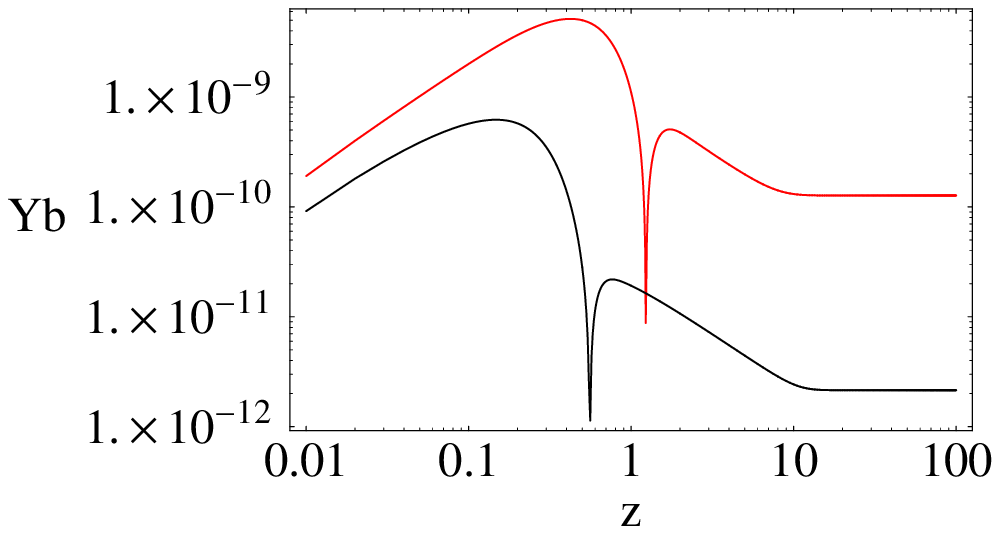}\hspace{1cm}\vspace{-1cm}
\includegraphics[width=6cm]{M1m1-2}
\caption{{\it Left:} Baryon asymmetry for specific lepton mass matrices
including flavour effects (upper) and without flavour effect (lower). From 
Abada et al. \cite{adx06}. {\it Right:} Domains of the ($M_1$-$m_1$) plane 
with different relevance of flavour effects; the two thick solid lines
border the region where successful leptogenesis is possible. From \cite{bdr06}.
}
\end{figure}

An important recent development in the theory of leptogenesis concerns the
effect of the flavour composition of heavy neutrino decays on the generated
lepton asymmetry \cite{nnx06}. Particularly interesting is the possible
connection between the baryon asymmetry and $C\!P$ violation at low energies
\cite{adx06}. Flavour effects can significantly enhance the generated
baryon asymmetry (cf.~Fig.~8) and therefore relax the upper bound on the light 
neutrino masses given in (\ref{win}). To quantify this effect which strongly
depends on the neutrino mass parameters (cf.~Fig.~8), a full quantum 
kinetic description of the leptogenesis process is required \cite{bdr06}.
Several groups have started to study leptogenesis on the basis of 
Kadanoff-Baym equations \cite{bf00}.

Over the years much work has also be done on the connection between 
leptogenesis and neutrino mass matrices which can account for low-energy
neutrino data. Many interesting models, some also very different
from the scenario considered above, have been discussed in the literature 
\cite{models}. Of particular interest is the connection with $C\!P$ violation 
in other low energy processes \cite{br05}. Together with leptogenesis, improved
measurements of neutrino parameters will have strong implications
for the structure of grand unified theories.

An alternative to thermal leptogenesis is nonthermal leptogenesis \cite{bpy05}
where the heavy Majorana neutrinos are not produced by thermal processes. 
These models are less predictive but arise naturally in many extensions of
the standard model. 

An intriguing aspect of thermal leptogenesis is its incompatibility with
the most popular supersymmetric extensions of the standard model where
the lightest neutralino is the dominant component of cold dark matter
and a heavy gravitino, decaying after nucleosynthesis, requires a reheating
temperature in the early universe much below the temperature needed for
leptogenesis. This clash has triggered much work on alternatives to WIMP dark 
matter. An attractive possibility is gravitino dark matter 
(cf.~\cite{bpy05,ps06,bcx07}) which can have striking effects at the LHC as
well as in gamma-ray astronomy.
  
\section{Conclusions}

40 years after Sakharov's work on the cosmological matter-antimatter asymmetry
we have several viable models of baryogenesis, the most predictive ones
being electroweak baryogenesis and leptogenesis. In fact, based on our
theoretical understanding of the electroweak phase diagram, electroweak 
baryogenesis in the standard model has already been excluded by the LEP bound
on the Higgs mass. Supersymmetric electroweak baryogenesis will soon be
tested at the LHC.

Detailed studies of the nonequilibrium leptogenesis process have led to
the preferred neutrino mass window $10^{-3}\ {\rm eV} < m_i < 0.1\ {\rm eV}$ in
the simplest scenario with hierarchical heavy neutrinos. The consistency with 
the experimental evidence for neutrino masses has dramatically increased the 
popularity of the leptogenesis mechanism. It is exciting that new experiments 
and cosmological observations will probe the absolute neutrino mass scale in
the coming years. However, more work is needed on the full quantum mechanical
treatment of leptogenesis, in particular the flavour dependence.

All baryogenesis mechanisms are closely related to the nature of dark matter.
A discovery of the standard supergravity scenario at LHC could be consistent
with electroweak baryogenesis but would rule out the simplest version of
thermal leptogenesis. On the other hand, evidence for gravitino dark matter
can be consistent with leptogenesis. Finally, the discovery of macroscopic 
dark matter like Q-balls would point towards nonperturbative dynamics of scalar
fields in the early universe and therefore favour Affleck-Dine baryogenesis.





\bibliographystyle{aipproc}   


\hyphenation{Post-Script Sprin-ger}

\end{document}